\documentclass[aps,prb,twocolumn,showpacs,preprintnumbers,amsmath,amssymb]{revtex4}
\input{epsf}
\input epsf.sty
\input psfig.sty

\begin{document}

\title{Metastable states in the planar $2d$ $XY$ model and dissipation
in superfluid flow}

\author{G.G. Batrouni}
\affiliation{Department of Physics, NTNU, N--7491 Trondheim, Norway}
\affiliation{Institut Non-Lin\'eaire de Nice, Universit\'e de Nice--Sophia
Antipolis, 1361 route des Lucioles, 06560 Valbonne,
France\footnote{Permanent address.}}

\begin{abstract}
We use the Metropolis algorithm to study the stability of superfluid
flow in a model system, namely the two-dimensional planar $XY$
model. Flow properties are examined by studying the behaviour of the
system in meta-stable ``twisted'' states. We demonstrate the stability
of superfluidity in this model and we discuss the Meissner effect and
velocity quantization. We also study the critical velocity and
dissipation by vortex creation and rotational flow and their
dependence on the geometry of the system. An expression for the
average superfluid velocity as a function of time, ${\bar v}_s(t)$, is
obtained and compared with experimental results.
\end{abstract}
 

\maketitle

\section{Introduction}

Interest in superfluidity has never waned. Quite the contrary, since
the experimental realization of trapped atomic Bose-Einstein
condensates \cite{ketterle,anderson} (BEC) activity in this field has
intensified dramatically and the recent apparent discovery of a
supersolid phase in solid helium\cite{sshelium} will most likely
increase the level of interest. In addition, there is intense interest
in the low temperature properties of bosons where many questions
remain concerning the phase diagrams of model systems and the phase
transitions from the superfluid phase to various exotic phases.

An important question concerning superfluids is that of dissipation
and the critical velocity. Perhaps one of the first things one thinks
of at the mention of superfluidity is ``flow without friction''. Very
early on, Landau~\cite{landau1} realized that the superfluid can be
treated as a dilute gas of noninteracting quasi-particle excitations
(phonons) and obtained the famous and often cited Landau stability
criterion for superfluids~\cite{landau2}:

\begin{equation}
\label{vcritlandau}
v_c^L = min_k[\omega(k)/k].
\end{equation}
The dispersion relation, $\omega(k)$, is the energy of a
quasi-particle excitation of momentum $\hbar k$ in the fluid, and
$v_c^L$ is the {\it Landau critical velocity} above which dissipation
sets in. Clearly, if the dispersion relation $\omega(k)$ is linear, at
least for small $k$, the Landau critical velocity is finite. On the
other hand, $v_c^L=0$ for quadratic dependence of $\omega(k)$ on
$k$. When $v_c^L \neq 0$, dissipationless flow at finite velocity is
possible and the superfluid is said to be ``stable''.

To measure the critical velocity of a superfluid such as liquid
$^4$Helium at a temperature $T<T_{\lambda}=2.18$K, one might, for
example, pull an object through the superfluid: The velocity at which
a drag force starts to act on this object can be thought of as a
critical velocity. Such experiments have been done~\cite{dragion} and
find a critical velocity which depends on the superfluid density,
$\rho_s$, but not on the geometry of the system. Furthermore, this
critical velocity was found to be of the order of a few tens of
meters/second which is in good agreement with the Landau prediction,
Eq.~(\ref{vcritlandau}), when applied to Helium.

Another way to measure the critical velocity is to set the superfluid
in motion in a tube (or through an orifice) and measure the superflow
velocity at which dissipation sets in. This is perhaps more appealing
physically than the first experiment because it involves a flowing
superfluid and because one may make connections with persistent
currents observed in superconducting systems. Such flow experiments
have also been performed~\cite{vcrit3d} and find that the critical
velocity is of the order cm/s or less and {\it does} depend on the
size of the orifice! In addition, it was found that the higher
critical velocities are obtained with the smaller orifices which is
perhaps counterintuitive. These results are surprising when compared
with the Landau prediction. It is thus clear experimentally that these
``critical'' velocities are not equivalent.

The widespread application of the Landau criterion as a test of
stability, Eq.~(\ref{vcritlandau}), has been
criticized~\cite{leggett1,leggett2} because it applies to the case of
an object moving in the fluid and not to the case of the fluid itself
flowing through orifices or in a torus. Perhaps a reason for the
ubiquity of the Landau criterion is that in many cases one can
calculate, if only approximately, the dispersion relation
$\omega(k)$. It is much more difficult to calculate, even numerically,
the critical velocity observed in flow experiments.

In this paper, we shall adopt the viewpoint of
reference~\cite{leggett2} and after a brief review, we shall present
our simulation. As usual, we have in mind here the two-fluid model of
superfluids. In the superfluid phase, the total density $\rho$ is the
sum of two densities,

\begin{equation}
\label{2fluid}
\rho = \rho_n + \rho_s,
\end{equation}
where $\rho_n$ ($\rho_s$) is the normal (super) fluid density.  The
superfluid component does not carry entropy and flows without
dissipation. The total particle current is then

\begin{equation}
\label{current}
{\vec j} = \rho_s {\vec v}_s + \rho_n {\vec v}_n,
\end{equation}
where ${\vec v}_s$ (${\vec v}_n)$ is the velocity of the superfluid
(normal) component. To obtain the expression for ${\vec v}_s$, we
write the superfluid wavefunction as
\begin{equation}
\label{wavefct}
\psi({\vec r})= {\rm e}^{i\theta({\vec r})} \psi_0(\vec r),
\end{equation}
where $\psi_0(\vec r)$ is real. Then
\begin{equation}
{\vec v}_s = \langle \psi| {{\hat p}\over m} |\psi \rangle
= {\hbar \over m}\langle \nabla \theta \rangle,
\label{Vs}
\end{equation}
where ${\hat p}$ is the momentum operator. The superfluid velocity is
proportional to the average gradient of the phase of the wavefunction:
Clearly, if the phase does not change coherently throughout the volume
of the system, ${\vec v}_s=0$.

Following reference~\cite{leggett2}, we consider two situations which
demonstrate fundamental defining properties of the superfluid.

(1) A torus containing liquid $^4He$ at $T>T_{\lambda}$ is spun around
its axis at very low angular velocity~\cite{lowangmom}. Eventually the
liquid will come to equilibrium with the moving walls. Reducing $T$
below $T_{\lambda}$, the liquid goes into its superfluid phase and the
superfluid component is observed to come to rest and, to conserve
angular momentum, the torus and normal fluid gain angular
momentum. The experiment~\cite{hess} demonstrates the analogue of the
{\it Meissner effect} in superconductors.

(2) Starting with the same setup as above, the torus is spun at high
angular velocity~\cite{lowangmom}. The temperature is then reduced
below $T_{\lambda}$ and the torus brought to rest. Eventually the
normal component will itself reach equilibrium with the walls and come
to rest. It can then be verified that the angular momentum of the
stationary torus is non-zero: The superfluid component is still
flowing and may continue to do so for a very long time. This is the
phenomenon of persistent dissipationless flow similar to persistent
currents in superconductors. Such experiments have been done by
several groups, see for example references~\cite{persistflow,eckholm}.

These two fundamental properties may be understood with the aid of the
velocity quantization condition first proposed by
Onsager~\cite{onsager1}
\begin{equation}
\oint {\vec v}_s.{\rm d}{\vec l} = n \kappa_0,
\label{velocity}
\end{equation}
where $\kappa_0=h/m$ is the flux quantum, $h$ is Planck's constant and
$m$ the particle mass. Clearly, the closed integration path must
enclose a ``hole'' in the system, either a vortex or a physical hole,
otherwise the path can be shrunk continuously to a point and only
$n=0$ survives. Another important property of superflow, which will
come into play below, is that it is irrotational,
\begin{equation}
\nabla \times {\vec v}_s = 0.
\label{irrotation}
\end{equation}

\begin{figure}
\psfig{file=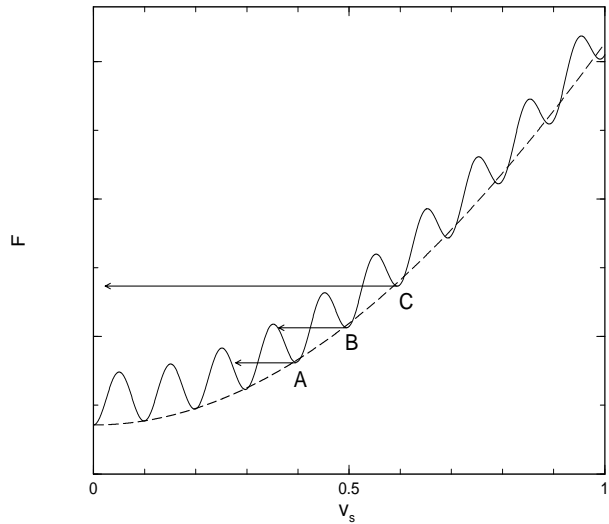,height=7.00cm,width=8cm,angle=-90}
\caption{Qualitative form of the free energy (arbitrary units)
versus velocity in the presence of superfluidity for a $64\times 64$
system. When the velocity satisfies Eq.~(\ref{velocity}), the free
energy has a local minimum (for the values of $v_s$ shown) which
explains the metastability of persistent superfluid flow and the
Meissner effect. As $v_s$ increases, the corresponding local minima
get shallower and eventually disappear rendering the flow unstable.}
\label{FvsVel}
\end{figure}

Equation~(\ref{velocity}) demands that persistent flow take place at a
well defined value $n\kappa_0$. This is not an equilibrium situation
since the superfluid can reduce its free energy by coming to rest. The
persistence of the flow, at least for $v_s$ below a critical value,
means that the fluid is in a {\it metastable} state which will
eventually decay into the equilibrium stable state at
rest~\cite{leggett2,hohenberg}. As we shall see, the lifetime of this
metastable state depends on the velocity itself and on
$T$. Figure~\ref{FvsVel} shows qualitatively the form of the free
energy as a function of velocity~\cite{leggett2,hohenberg}. This also
makes clear that transitions between local minima should be possible
and lead to the loss of superfluid momentum, in other words the
begining of dissipation due to excessive velocity. Such phase slips
were indeed observed experimentally~\cite{packard}. We now make
precise what was meant by ``low'' and ``high'' velocities in the
discussion of the two experiments above. The Meissner effect takes
place when the initial velocity corresponds to less than half a
quantum and the superfluid, seeking the nearest velocity satisfying
Eq.~(\ref{velocity}), comes to rest, {\it i.e.} excludes all
flux. When the initial velocity is high, {\it i.e.} larger than half a
quantum, the superfluid will seek the nearest velocity satisfying
Eq.~(\ref{velocity}) and settle into the corresponding metastable
state.

It was suggested~\cite{onsager1,feynman1} that vortex formation is
behind these transitions and although this mechanism is generally
accepted, the details are still not well understood and no numerical
simulations have been done.

In the following sections we shall address these questions numerically
for a model system.  The paper is organized as follows. In section
{\bf II} we briefly review the two dimensional $XY$ model. In section
{\bf III} we present our results for the non-equilibrium simulations
including velocity quantization, superfluid density, transitions and
flux dissipation, scaling of lifetimes with the geometry of the system
and comparisons with experiments. Conclusions are in section {\bf IV}.

\section{Model}

Addressing these questions with Quantum Monte Carlo simulations of
$^4He$ or of model systems such as the bosonic Hubbard model, poses
very difficult algorithmic problems. As discussed above, putting
the superfluid in motion requires a gradient in the phase of the
wavefunction. This introduces a complex phase in the Boltzmann weight
rendering the simulation extremely difficult.

Instead, we shall use classical Monte Carlo to simulate the two
dimensional planar $XY$ model which has been studied extensively in
connection with two dimensional superfluidity. This model exhibits the
well known Kosterlitz-Thouless ($KT$) transition at $T_{KT}$.  For
$T>T_{KT}$, there is a condensation of dissociated vortices, the
phase is disordered and the correlation function decays exponentially.
For $T<T_{KT}$ it has a spin-wave phase characterized by tight binding
of vortex-antivortex pairs and power law correlation function. There
is no symmetry breaking in this phase and, therefore, no
magnetization. In the language of superfluids, the absence of
magnetization is equivalent to the absence of BEC. In addition, the
low energy excitations, the spin waves, have a quadratic dispersion
relation: $\omega({\vec k}) \sim {\vec k}^2$. On the face of it, this
might be taken to imply an unstable superfluid according to the Landau
criterion. However, it must be recalled the these spin waves are
thermodynamic in nature, not dynamic: This model has no intrinsic
dynamics, unlike the three component $XY$ model which has been shown
to posses dynamic spin wave excitations with linear
dispersion\cite{evertz}. 

One may justify using this model to study the dissipation properties
of two dimensional superfluids by recalling that experiments on $^4$He
films~\cite{filmexp} show that the transition to the superfluid phase
is indeed in the $KT$ universality class and that these films have
finite critical velocity~\cite{filmvcrit,eckholm}. In addition, most
explanations of dissipation in superfluid
flow~\cite{iord,langfisher,huberman,halperin} are based on vortex
formation which is certainly a defining property of the
two-dimensional $XY$ model.  It is, therefore, reasonable to use this
model to address the questions of surperfluid stability and
dissipation. The quadratic dispersion of the low-lying excitations and
the absence of BEC make these questions even more interesting.

Consider, therefore, a two dimensional square lattice with
$N=L_x\times L_y$ sites. The partition function of the $XY$ model is
then given by
\begin{equation}
Z= \int_{-\pi}^{+\pi} \prod_{i=1}^N d\theta_i\, {\rm exp} \Bigl(\beta
\sum_{\langle ij\rangle}{\rm cos}(\theta_i-\theta_j)\Bigr),
\label{zxy}
\end{equation}
where $\langle ij\rangle$ denotes nearest neighbour sites and
$\beta=1/kT$ with $k$ the Boltzmann constant.  Since the superfluid
component does not carry any entropy, the increase in free energy when
the superfluid is in motion is due only to its kinetic energy. Then,
with $\rho_s$ the superfluid particle density and taking ${\vec v}_s$
purely in the $x$ direction, one may write for small $v_s$,

\begin{equation}
\label{freev}
F(v_s) \approx F_0 +{1\over 2} L_xL_y\rho_s m v_s^2 
\end{equation}
from which immediately follow the expressions for the superfluid
momentum density,
\begin{equation}
\label{supermom}
m \rho_s v_s = {1\over {L_xL_y}} 
{{\partial F(v_s)}\over {\partial v_s}},
\end{equation}
and the superfluid particle density
\begin{equation}
\label{superden}
\rho_s = {1\over {L_xL_y}}{1\over m} {{\partial^2 F(v_s)}\over
{\partial v_s^2}}.
\end{equation}

It is clear from Eq.~(\ref{velocity}) that the lowest velocity must
correspond to $n=1$ and consequently, for $v_s$ to be small enough to
justify Eq.~(\ref{freev}), the system must be large. Equation
(\ref{superden}) is often expressed in words by saying that the
superfluid density is the curvature of the free energy as a function
of the superfluid velocity. That this is not quite true is clear from
Eq.~(\ref{velocity}) and Fig.~\ref{FvsVel} which together emphasize
that $\rho_s$ is the curvature of the parabola passing through the
first few local minima of the free energy as a function of
$v_s$. However, for brevity, we shall refer to this as the curvature
of the free energy.

Recalling that $F=-{\rm ln}Z/\beta$, it is easy to apply
Eqs.~(\ref{supermom},\ref{superden}) to the $XY$ model and obtain
\begin{equation}
\label{supermomxy2}
{{\partial F}\over {\partial v_s}} = \Bigl \langle
\sum_{\langle i,j \rangle:x}{\rm sin}(\theta_i-\theta_j)\Bigr \rangle
\end{equation}
\begin{eqnarray}
\nonumber
{{\partial\,^2 F}\over {\partial v_s^2}}&=& \beta \Biggl \{
\Bigl\langle \sum_{\langle i,j \rangle:x} {\rm sin}(\theta_i-\theta_j)
\Bigr\rangle^2 \\
\nonumber
&&- \Bigl\langle \bigl (\sum_{\langle i,j \rangle:x} {\rm
sin}(\theta_i-\theta_j) \bigr )^2 \Bigr\rangle \Biggr \}\\
\label{superdenxy}
&&+ \Bigl\langle \sum_{\langle i,j \rangle:x} {\rm cos}(\theta_i-\theta_j)
\Bigr\rangle 
\end{eqnarray}
where the notation $\langle i,j \rangle$:$x$ means the sum is
performed over nearest neighbours only in the $x$ direction (since we
took $v_s$ to be in that direction). Equation~(\ref{supermomxy2})
allows the determination of $\rho_s$ in a {\it nonequilibrium flow
situation}, while Eq.~(\ref{superdenxy}) gives $\rho_s$ also at
equilibrium where $v_s=0$.

\section{Non-equilibrium simulations}

\subsection{Metastability and superfluid density}

In this section we study some of the non-equilibrium properties of the
two-dimensional $XY$ model. In particular, we are interested in the
transitions among the local minima of Fig.~\ref{FvsVel}, in other
words the onset of dissipation in the superflow and its relation to
the critical velocity and thermodynamic stability of the
superfluid. The transitions between the local minima are driven by
thermal fluctuations as the system attempts to minimize its free
energy.  Different simulation algorithms will lead to different
lifetimes of the metastable states. However, the scaling of these
lifetimes with the geometry of the system and the velocity of the
superfluid will be the same.We chose to use the single spin flip local
Metropolis algorithm. We shall see below that agreement with
experiments is very good.

\begin{figure}
\psfig{file=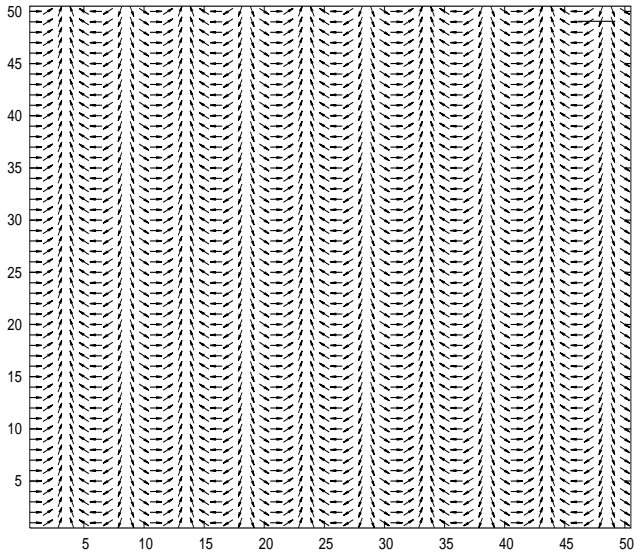,height=3in,width=3.5in,angle=-90}
\caption{Configuration with $v_s=n\kappa_0,\,\,n=5$ on a $50\times 50$
lattice. This configuration satisfies the velocity quantization
condition and locally minimizes the free energy for $n=5$ at
$T=0$. The direction of the arrows gives the value of $\theta_i$.}
\label{vs5}
\end{figure}

Non-equilibrium simulations are performed by placing the system in
initial configurations corresponding to flow at a chosen initial
superfluid velocity. Equation (\ref{Vs}) shows that the superfluid
flows if there is a phase gradient, while Eq.~(\ref{velocity}) places
a condition on the allowed values of $v_s$. Figure \ref{vs5} shows an
initial configuration corresponding to a flowing superfluid with flux
$n\kappa_0=5\kappa_0$. It is also clear that this configuration
corresponds to irrotational flow.

\begin{figure}
\psfig{file=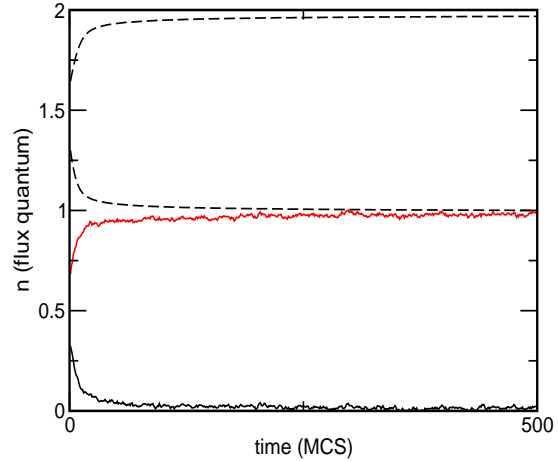,height=3in,width=3.5in,angle=-90}
\caption{The flux quantum, $n$ as a function of time in Monte Carlo
Sweeps (MCS) for $100\times 100$ system at $T=0.25< T_{KT}$. We see
that $n$ flows to the nearest integer value clearly exhibiting the
Meissner effect and velocity quantization.}
\label{meissner}
\end{figure}

We first show that superfluid flow in the two dimensional $XY$ model
is (meta)stable by demonstrating the existence of the Meissner effect
and velocity quantization.  As discussed in the introduction, if the
initial $n$ is less the $1/2$, a ``stable'' superfluid will satisfy
velocity quantization by coming to rest thus exhibiting the Meissner
effect. On the other hand, if $n>1/2$, the system will evolve to the
nearest integer value of $n$ changing its velocity in the process.
Figure~\ref{meissner} shows simulation results for four initial values
of the flux quantum, $n=0.4, 0.6, 1.4, 1.6$ for a $100\times 100$
system at $T=0.25<T_{KT}$. The behaviour of single configurations is
shown for $n=0.4, 0.6$ while for $n=1.4$ and $n=1.6$ we show averages
over $1000$ configurations (the dashed lines). The figure shows
clearly that the Meissner effect and velocity quantization are both
present in the two dimensional $XY$ model below $T_{KT}$. This means
that the free energy does indeed have the form depicted in
Fig.~\ref{FvsVel} otherwise there would be no reason for the flux
quantum to get stuck at integer values as shown in
Fig.~\ref{meissner}.

\begin{figure}
\psfig{file=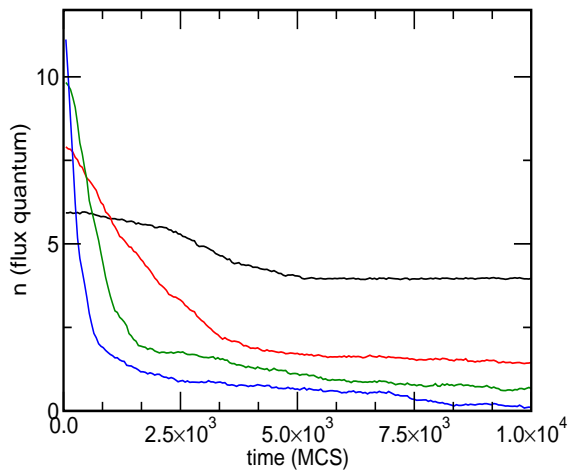,height=3in,width=3.5in,angle=-90}
\caption{The flux quantum $n$ versus MCS for $100\times 100$ system at
$T=0.5$. It is clear that the final value of $n$ depends on the
initial value: It is not necessary for the system always to tunnel to
the $n=0$ state, nor does it have to tunnel from $n$ to $n-1$. The
initial values for the flux are $n=6, 8, 10, 12$.}
\label{tunnelflux}
\end{figure}

That integer flux configurations are metastable, rather than stable,
is shown in Fig.~\ref{tunnelflux} where transitions are seen. It is
also clear from this figure that transitions do not always change the
flux by the same amount. The change in the flux, and consequently
dissipation and the final metastable or stable configuration, depend
on the starting value: The larger the initial flux, the larger the
change and, consequently, the smaller the value of the final flux. The
reason for this will be discussed below where we will study these
transitions in more detail.  Such transitions between local minima,
also referred to as phase slips, have been observed
experimentally~\cite{packard}.

\begin{figure}
\psfig{file=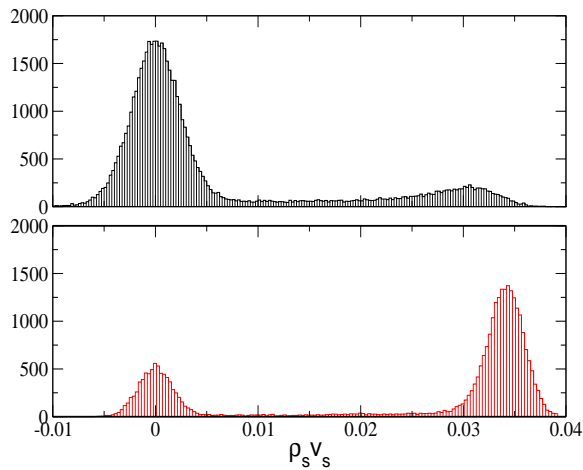,height=3in,width=3.5in,angle=-90}
\caption{Histogram of the superfluid momentum for a system of size
  $128\times 128$. Upper panel: $T=0.9$ and $45$ simulations. Lower
  panel: $T=0.825$ and $30$ simulations. See discussion in text.}
\label{histog}
\end{figure}

We now compare the superfluid density using equilibrium and
nonequilibrium measurements. The equilibrium measurements were done,
as usual, by using Eq.~(\ref{superdenxy}) and random initial
configurations with $n=0$. The nonequilibrium measurements were
performed by putting the system in an $n=1$ initial configuration and
using Eq.~(\ref{supermom}) and~(\ref{supermomxy2}) to measure the
superfluid density. For $T<0.7$ the $n=1$ metastable state is so long
lived that even after ten million sweeps the transition does not occur
(for $128\times 128$ system). For such long lived states the
measurements are simply performed as if the system is in an
equilibrium configuration. However, as $T_{KT}$ is approached, the
lifetime of the metastable state becomes very short and more care must
be taken. In this temperature range we measure the superfluid density
as follows. The system is put in the $n=1$ initial configuration and
its evolution is followed until it makes the transition to the $n=0$
configuration, performing measurements every few sweeps. This is done
many times for the same temperature ($40$ to $300$ times); of course
the transition time can be very different from one simulation to
another at the same $T$ (this will be discussed below). The superfluid
momentum density\cite{footnotevs}, $\rho_s v_s$, from all the runs for
the same temperature is then histogrammed to decide if it is even
meaningful to calculate an average of this quantity. The results for
two temperatures are shown in Fig.~\ref{histog}. The relative heights
of the peaks is not important since the height of the peak at $\rho_s
v_s=0$ depends on how long the simulation is allowed to run after the
transition has taken place. What is important is the height of the
peak at nonzero momentum compared to its width and also compared to
the height of the histogram in the transition region between the two
peaks. It is clear that for the upper panel of Fig.~\ref{histog}, it
is meaningless to calculate the superfluid momentum in the metastable
state, while for the lower panel this quantity is well defined.

\begin{figure}
\psfig{file=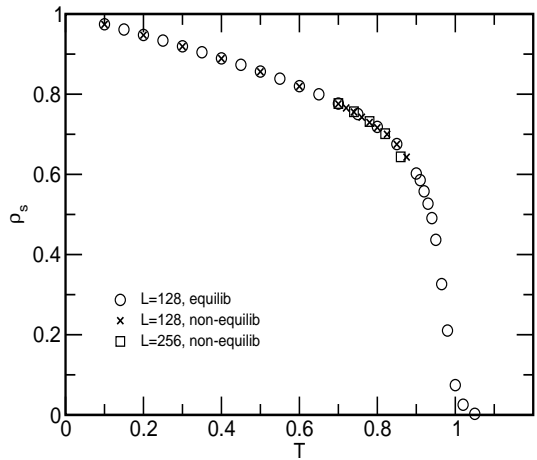,height=3in,width=3.5in,angle=-90}
\caption{The superfluid density, $\rho_s$, from equilibrium and
non-equilibrium measurements versus $T$. The error bars are smaller
than the symbols.}
\label{rhosvsT}
\end{figure}

With the help of such analysis we calculate $\rho_s$ in the
nonequilibrium situation and compare with the equilibrium values in
Fig.~\ref{rhosvsT}. The agreement is excellent which demonstrates
numerically the correctness of this approach. For $T>0.875$, we can no
longer extract $\rho_s$ this way: The initial superfluid velocity,
$v_s=2\pi/L_x$ (in units of $\hbar/m$ and for $L_x=128$) is greater
than the critical velocity at these temperatures and the superfluid
kinetic energy is quickly dissipated. In order to get closer to
$T_{KT}$, the initial $v_s$ must be smaller which is impossible for
this system size since already $n=1$. To achieve lower $v_s$, a larger
system would be needed; this is another demonstration of the important
interplay between system size and $v_s$.

Note that the excellent agreement between the equilibrium and
non-equilibrium measurements of $\rho_s$ (which we have also verified
with simulations at higher velocities) does not support Eq.(11) of
reference \cite{huberman} which claims a strong dependence of $\rho_s$
on $v_s$.

An interesting aspect of the nonequilibrium measurements of $\rho_s$ is
the possibility of getting direct evidence for the universal jump
condition~\cite{jump},
\begin{equation}
\label{univjump}
\rho_s(T_{KT}) = {2\over \pi} T_{KT}.
\end{equation}
In~\cite{minnhagen} a method based on higher order derivatives of the
free energy was presented as a way for direct detection of this
jump. In the present method, it is clear that when the system is close
enough to $T_{KT}$ the superfluid momentum will simply vanish
discontinuously. Already for $L=128$, this happens for $T$ just above
$0.875$ (the last nonequilibrium point shown in Fig.~\ref{rhosvsT}),
whereas $T_{KT}\approx 0.9$.

\medskip

\subsection{Vortices and dissipation}

We now turn to the dissipation mechanism, {\it i.e.} the excitations
which take the system from one local minimum to another.
Onsager~\cite{onsager1} and Feynman~\cite{feynman1} argued that the
creation of large vortices in the flowing superfluid was responsible
for these dissipative transitions. Although the general features of
this idea are widely accepted, disagreement remains on the
details. Here we shall study and confirm numerically that, in the two
dimensional $XY$ model, transitions between the metastable persistent
flow states proceeds via the formation of large vortex-antivortex
pairs oriented orthogonally to the direction of flow. We will describe
how it happens and study its dependence on the width of the system and
the flux velocity.

\begin{figure}
\psfig{file=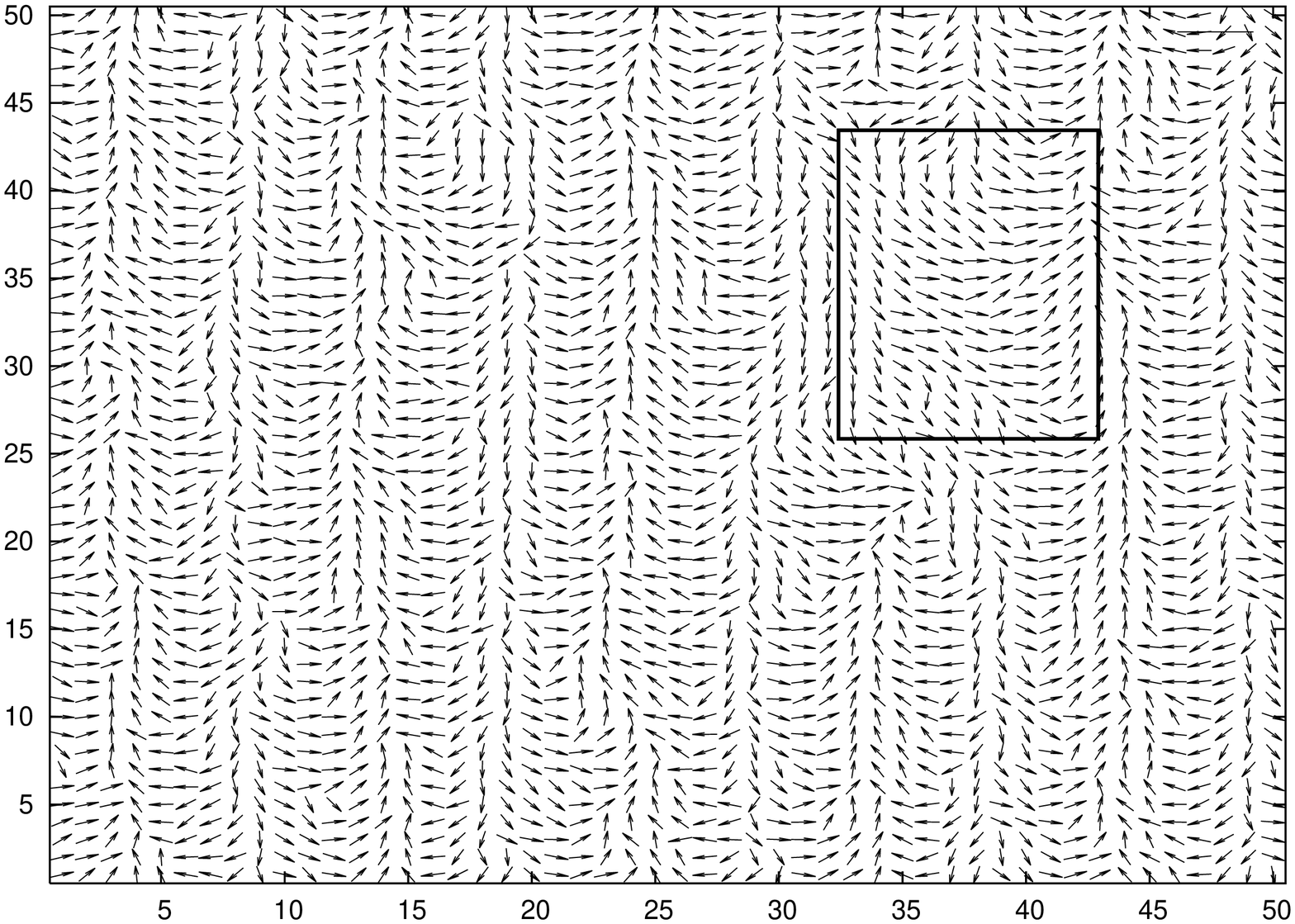,height=3in,width=3.5in}
\caption{The configuration in Fig.~\ref{vs5} after $1100$ MCS at
  $T=1/3$. The square shows the region where two bands of spins
  differing by $2\pi$ have merged. A vortex (antivortex) can be seen
  below (above) the square.}
\label{zipper}
\end{figure}

Our numerical simulations demonstrate that, for $T<T_{KT}$, thermal
fluctuations take place in such a way that the large scale band
structures, Fig.~\ref{vs5}, are maintained due to the spin
stiffness. As individual spins undergo thermal fluctuations, the bands
of approximately parallel spins fluctuate as large scale elastic
objects maintaining their large scale form and spanning the system (in
the $y$ direction by choice). Deformations of these elastic objects
cost energy proportional to the curvature of the deformation and to
the spin stiffness, {\it i.e.} $\rho_s$. Eventually, two bands whose
spins differ by $2\pi$ are deformed enough to touch. When this
happens, the $2\pi$ difference between these bands loses its meaning
where they touch, but away from the contact region the difference
still exists. This topological ambiguity is in fact the
vortex-antivortex excitation which can trigger the transition. This
situation is shown in Fig.~\ref{zipper} which is a snapshot of the
system depicted in Fig.~\ref{vs5} at $T=1/3$ after $1100$ Monte Carlo
sweeps. The square in Fig.~\ref{zipper} shows where two bands of spin
differing by $2\pi$ have touched. A vortex is visible just below the
square and an antivortex just above it; also visible is a flux line
going down from the vortex and connecting it to the antivortex (we
have periodic boundary conditions). In two dimensions, for $T<T_{KT}$,
the energy of such excitations is of the order of ${\rm ln}r$, where
$r\approx L_y$ is the separation of the vortex-antivortex pair.

When such contact is made, it may be very difficult to break and
dissipation of a flux quantum can then proceed by zipping together the
two bands into a single one thus eliminating a quantum of flux. The
reason for this is that for $T<T_{KT}$ the vortices are confined: It
is favorable for a vortex to find an antivortex and annihilate. In the
situation of Fig.~\ref{zipper}, the logarithmic confining potential
will pull them together with the vortex moving down and the antivortex
moving up to meet and annihilate zipping the the two bands in their
wake. In the case of non-periodic boundary conditions in the $y$
direction, the vortex and antivortex will each be attracted to its
image thus moving outward and crashing against the wall. The outcome,
in both cases, is that the two bands are zipped together and a quantum
of flux disappears, {\it i.e.}  dissipates.

Note that this mechanism, while involving vortices, is not quite the
same as that discussed in reference~\cite{langfisher}.
In~\cite{langfisher}, the vortex is assumed to nucleate at a singular
point where the wavefunction vanishes, $|\psi|\to 0$, which implies a
vanishing density at that point. In the mechanism discussed here, only
the phase, modeled by the angles $\theta$, fluctuates: The
transitions here are caused by phase, not density, fluctuations.

 whereby a single
band snaps into two parts each of which subsequently shrinks and melts
away. While we have observed such events in our simulations, it
appears that in the temperature and flux quantum ranges we studied
that the zipper mechanism dominates.

The role of the velocity is now straightforward to describe
qualitatively.  The higher the velocity, the more closely packed the
bands become and thus the higher the likelihood that more than one
contact region be established. Consequently, several vortex-antivortex
pairs may be created with the result that the transition can dissipate
more than one flux quantum. This explains what is observed in
Fig.~\ref{tunnelflux} where it is seen that more flux quanta are
dissipated if the initial velocity is higher. If the velocity is very
high, so many vortices are created that the transition is essentially
immediate.

It is interesting to note that whereas the superflow is in general
irrotational, this is no longer the case while the system is
undergoing a transition. This is clearly seen in Fig.~\ref{zipper}
where one observes sheer in the flow: The ${\hat x}$ flow velocity in
the zones where the two bands have not yet touched is larger than that
where contact has been established thus producing sheer at the
interface.

\begin{figure}[h]
\psfig{file=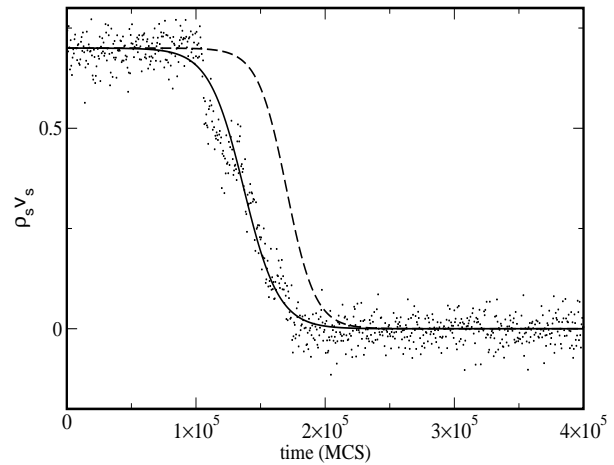,height=3in,width=3.5in,angle=-90}
\caption{The superfluid momentum versus time in Monte Carlo Steps for
  a $128\times 128$ system at $T=0.83$. The line is a fit of the form
  $\rho_s v_s=a(1-{\rm tanh}(b(t-\tau_{esc})))$. The dashed line is a
  fit to the transition of another configuration which is not shown
  for clarity.}
\label{tanh}
\end{figure}

Before discussing the scaling of the lifetime of the metastable
states, we first show that the escape time from a local minimum is
well defined and may be measured accurately. Figure~\ref{tanh} shows a
typical evolution of the superfluid momentum as a function of time in
Monte Carlo Steps (MCS). The solid line is a fit of the form $\rho_s
v_s=a(1-{\rm tanh}(b(t-\tau_{esc})))$ which allows us to measure the
lifetime, $\tau_{esc}$, of the initial metastable state. The dashed
line is a fit of the same form to the evolution of another
configuration which we do not show. We shall return to this figure in
the next subsection.

\begin{figure}[h]
\psfig{file=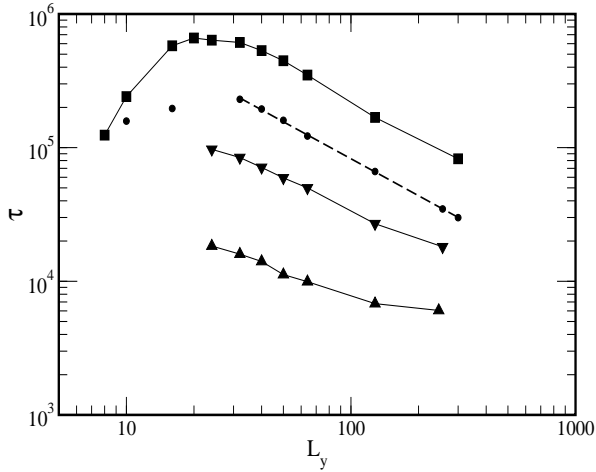,height=3in,width=3.5in,angle=-90}
\caption{The average lifetime $\tau$ (MCS) versus $L_y$ for
  $8\leq L_y \leq 300$ and $L_x=128$. The transitions are
  from $n=7$ to $n=6$ (up triangles), $n=6$ to $n=5$ (down triangles),
  $n=5$ to $n=4$ (squares). For these cases $T=0.5$.  Circles:
  transition from $n=12$ to $n=11$ at $T=0.25$. The averages are
  calculated over between $50$ and $10^{3}$ realizations. The dashed
  line is a fit giving $\tau=5.7\times 10^{6}/L_y^{0.9}$, the solid
  lines are to guide the eye.}
\label{powerlaw}
\end{figure}

As mentioned in the introduction, the critical velocity of superfluid
helium passing through orifices {\it decreases} as the opening size is
{\it increased}. Here we show that similar behavior is exhibited by
the two dimensional $XY$ model. In Fig.~\ref{powerlaw} we show as a
function of the width of the system, $L_y$, the average time (in MCS)
to make a transition from the $n$ to the $n-1$ state for $T=0.25,
0.5$. We see that, for $L_y\geq 32$, $\tau$ decreases as a power law
as the system gets wider. This decrease in $\tau$ with increasing
$L_y$ may be understood qualitatively with the help of the mechanism
shown in Fig.~\ref{zipper}: Increasing $L_y$ makes it easier, at
constant spin stiffness, to bend the spin bands by the needed amount
to make them touch since the {\it curvature} is smaller the larger the
$L_y$. However, a straightforward application of thermal activation
arguments fails to give the observed behaviour. The energy of a
vortex-antivortex pair in a the superfluid flow field is given
by~\cite{huberman}
\begin{equation}
E=2\pi\rho_s{\rm ln}L_y - (2\pi)^2 \rho_s n {L_y \over L_x},
\label{vortexenergy}
\end{equation}
where the distance between the vortex and antivortex is taken to be
$L_y$ as is seen in Fig.~\ref{zipper} and the superfluid velocity is
$2n\pi/L_x$. This gives a transition rate~\cite{eckholm}
\begin{equation}
r = \nu_0 L_y {\rm e}^{-\beta E},
\label{rate}
\end{equation}
where $\nu_0$ is the attempt rate. The average lifetime is then
\begin{equation}
\tau \sim L_y^{2\pi \rho_s\beta -1} {\rm e}^{-\beta \rho_s n (2\pi)^2
  L_y/L_x}.
\label{lifetime}
\end{equation}
This result which predicts an exponential decay of $\tau$ with $L_y$
does not agree with the observed numerical results~\cite{note}. 
For very narrow systems, $L_y=8, 10, 16$ in Fig.~\ref{powerlaw}, it is
seen that $\tau$ does not decrease with increasing $L_y$. This is
because the system is becoming one dimensional in which case the spins
are not stiff and therefore $\tau \to 0$.

\begin{figure}
\psfig{file=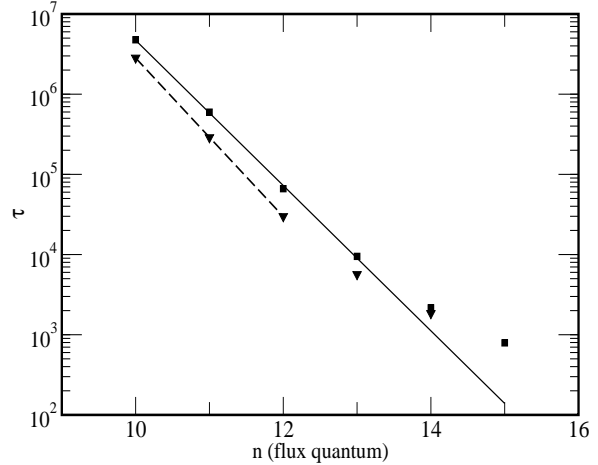,height=3in,width=3.5in,angle=-90}
\caption{The average lifetime of metastable states versus the flux
  quantum, ${\rm n}$, for $128\times 128$ (squares) and $128\times 130$
  (triangles) systems at $T=0.25$. The solid line is $\tau = 5.4\times
  10^{15}{\rm exp}(-2 {\rm n})$, the dashed line is $\tau = 1.9\times
  10^{16}{\rm exp}(-2.276 {\rm n})$.}
\label{exponential}
\end{figure}

Figure~\ref{exponential} shows the dependence of $\tau$ on $v_s=2\pi
n\kappa_0$. The metastable states ({\it i.e.} persistent superflow)
are shorter lived for larger velocities, $v_s$, which is in agreement
with experiments. The exponential decay with increasing velocity
appears in Eq.~\ref{lifetime} but with an exponent which is too large.
For the larger values of $n$, the decay is no longer exponential
because local minima become too shallow and also because the
simultaneous dissipation of flux quanta becomes more important.

\subsection{Comparison with experiments}

Decay of persistent currents has been studied both in bulk and
films. The results of references~\cite{langreppy,kojima,hallock}
were found to be consistent with time dependence of the superfluid
velocity in the form
\begin{equation}
v_s = A - B {\rm ln}t,
\label{vlog}
\end{equation}
where $A$ and $B$ are empirical constants. In other words, the
observed dissipation was very slow. However, reference~\cite{eckholm}
observed both slow and fast dissipation, the former well described by
Eq.(\ref{vlog}) while the latter much better described by 
\begin{equation}
v_s = {{A_1}\over {(1+B_1 t)^{r}}},
\label{veckholm}
\end{equation}
where $A_1,B_1$ and $r$ are empirical constants. Equation (\ref{vlog})
fails to describe the fast decays.

\begin{figure}
\psfig{file=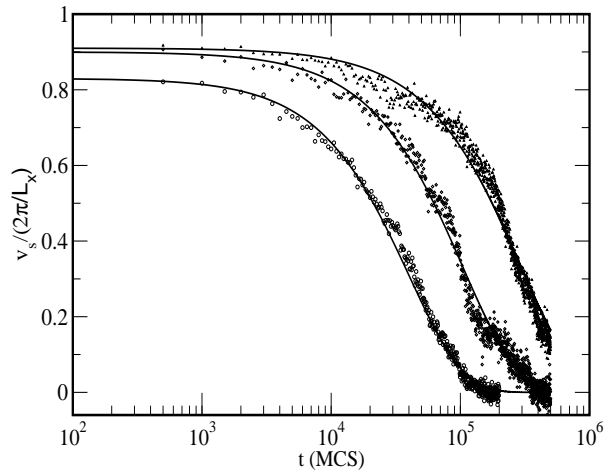,height=3in,width=3.5in,angle=-90}
\caption{Superfluid velocity versus time averaged over several
  realizations for a $128\times 128$ system. Lowest curve: $T=0.92$,
  $100$ realizations, middle curve: $T= 0.875$, $30$ realizations and
  upper curve: $T= 0.85$, $30$ realizations. See text after
  Eq.~(\ref{avevsf2}) for a discussion of the lines.}
\label{dissip}
\end{figure}

In this subsection we shall compare the dissipation in the $XY$ model
with these experimental results. Figure~\ref{tanh} shows, for one
configuration, the superfluid momentum as a function of time clearly
displaying the dissipation of kinetic energy. The behaviour in this
figure follows neither Eq.(\ref{vlog}) nor (\ref{veckholm}) and does
not resemble figures 6 or 7 of reference~\cite{eckholm} (see
Figs.~\ref{eckhomFig6} and~\ref{eckhomFig7} in this paper). However,
Fig.~\ref{tanh} shows the behaviour of only one configuration
dissipating exactly one flux quantum whereas in the experiments one
presumably observes an average of such processes. In other words, the
experimental situation represents many different regions of quantized
flux making transitions at different times. What is observed then is
the average dissipation as a function of time. To test this idea, we
performed simulations of the type shown in Fig.~\ref{tanh} but
averaging over many configurations. Three such averages are shown by
the points in Fig.~\ref{dissip}. This figure strongly resembles Fig. 6
of reference~\cite{eckholm} and the data are reasonably well described
by Eq.~(\ref{veckholm}) (not shown in the figure, the curves shown
will be discussed below). We therefore see that, at least for fast
dissipation, the $XY$ model is in very good qualitative agreement with
experiments adding support to the phase zipping dissipation mechanism
discussed above. Slow dissipation is harder to study numerically
because for long escape time it takes many more realizations to get
good statistics. However, the simulations we performed for this case
are also consistent with the experiments.

\begin{figure}
\psfig{file=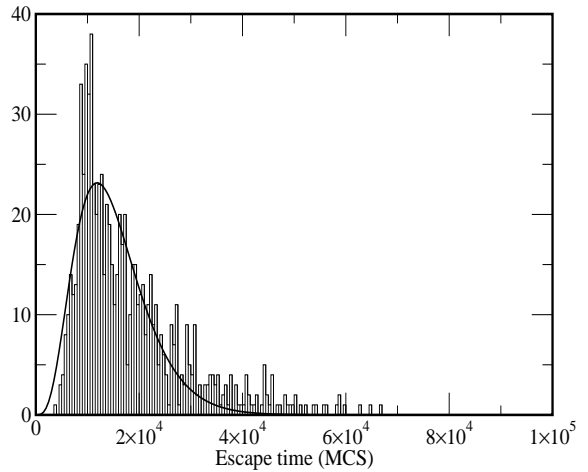,height=3in,width=3.5in,angle=-90}
\caption{The distribution of escape times from $n=6$ to $n=5$ for a
  $128\times 128$ system at $T=0.5$. The solid curve is a fit of the
  form Eq.~\ref{distau}.}
\label{taudistrib}
\end{figure}

\begin{figure}
\psfig{file=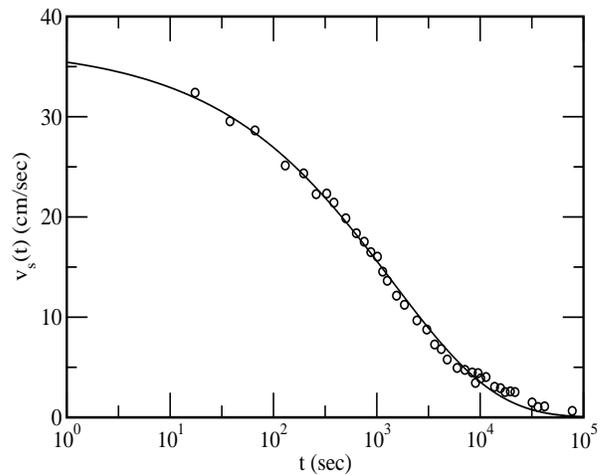,height=3in,width=3.5in,angle=-90}
\caption{Superfluid velocity versus time. The circles are the
  experimental data in figure 6 of reference~\cite{eckholm}. The solid
  curve is a fit using Eq.~(\ref{avevsf2}). The fitting parameters are
  $(\alpha=0,\gamma=0.043)$.}
\label{eckhomFig6}
\end{figure}

\begin{figure}
\psfig{file=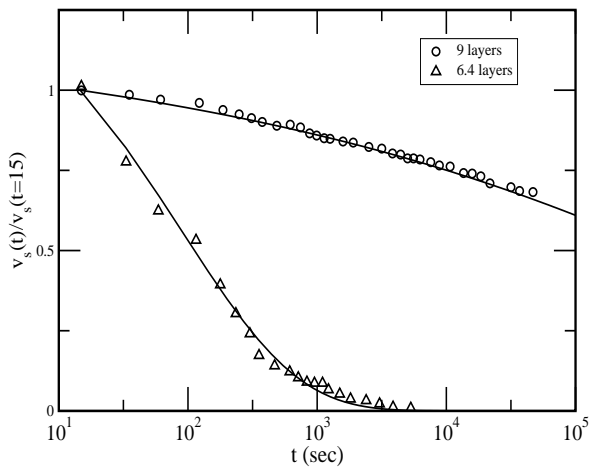,height=3in,width=3.5in,angle=-90}
\caption{Superfluid velocity versus time for films with different
  thickness at $T=1.45$K. The symbols are experimental data from
  figure 7 of reference~\cite{eckholm}. The solid curves are fits
  using Eq.~(\ref{avevsf2}). For $6.4$ layers
  ($\alpha=0.3,\gamma=0.5$) and for $9$ layers
  ($\alpha=-0.73,\gamma=0.001$)}
\label{eckhomFig7}
\end{figure}

To model the numerically observed time dependence of $v_s$ displayed
in Fig.~\ref{dissip}, one must examine the distribution of escape
times, $\tau_{esc}$, at a given $T$ and $v_s$. One such distribution
is shown in Fig.~\ref{taudistrib}. The solid curve is a fit of the
form
\begin{equation}
P(\tau_{esc}) = A \tau_{esc}^{\alpha}{\rm e}^{-\gamma\tau_{esc}},
\label{distau}
\end{equation}
where $A, \alpha$ and $\gamma$ are fitting parameters. For the case of
Fig.~\ref{taudistrib}, $\alpha=3.7$ and $\gamma=3.1\times
10^{-4}$. The average superfluid velocity, ${\bar v}_s$, is then given
by
\begin{equation}
{\bar v}_s(t) = \int_0^{\infty} {\rm d}\tau_{esc} v_s(t) P(\tau_{esc})
\label{avevsf}
\end{equation}
where $P(\tau_{esc})$ is normalized and where $v_s(t)$ may be taken of
the form discussed in Fig.~\ref{tanh}. However, to simplify the
discussion and allow exact integration of Eq.~(\ref{avevsf}), we may
take $v_s(t)=v_s(0)\Theta(\tau_{esc}-t)$ where $\Theta$ is the
Heaviside function. Equation~(\ref{avevsf}) then reduces to
\begin{eqnarray}
\nonumber
{\bar v}_s(t) &=& A \int_t^{\infty} {\rm
  d}\tau_{esc}\tau_{esc}^{\alpha}{\rm e}^{-\gamma\tau_{esc}} \\ 
& = & A^{\prime} \, \Gamma(1+\alpha,\gamma t)
\label{avevsf2}
\end{eqnarray}
where $\Gamma(1+\alpha,\gamma t)$ is the incomplete Gamma function,
$A^{\prime},\alpha$ and $\gamma$ are fitting parameters. The solid
curves in Fig.~\ref{dissip} are fits to the numerical data using
Eq.~(\ref{avevsf2}) with $(\alpha=0.08, \gamma=2.77\times 10^{-5})$
for the lowest curve, $(\alpha=0.04, \gamma=10^{-5})$ for the middle
curve and $(\alpha=0.02, \gamma=3.5\times 10^{-6})$ for the top
curve. We see that agreement with the numerical results is excellent.
As a further test of these ideas, we show in Fig.~\ref{eckhomFig6} a
fit to the experimental data in Fig. 6 of reference~\cite{eckholm}
using Eq.~(\ref{avevsf2}). The curve agrees remarkably well with the
experimental data, in fact much better than Eq.~(\ref{veckholm}), see
Fig. 6 in reference~\cite{eckholm}.

The escape time distribution, Eq.~\ref{distau}, used in this analysis
is reasonable for fast to medium dissipation as can be seen in
Fig.~\ref{taudistrib}. However, it is not clear that the same
distribution gives a reasonable description for extremely long lived
persistent flows of the type well modeled by Eq.~(\ref{vlog}). In
other words, the question is whether for very slow decays
Eq.~(\ref{avevsf2}) behaves like the experimental results,
i.e. roughly linearly in ${\rm ln}t$. To verify this, we show in
Fig.~\ref{eckhomFig7} fits of Eq.~(\ref{avevsf2}) to two data sets
from figure 7 of reference~\cite{eckholm}. We see that even for the
slowest dissipation reported, Eq.~(\ref{avevsf2}) gives very good
agreement with experiments. For the slow decay, though, it was
necessary to take $\alpha<0$ to get a good fit. This might cause
concern in view of Eq.~(\ref{distau}) which would then diverge for
$\tau_{esc}= 0$. However, negative $\alpha$ is needed only for the
very slow decays where $\tau_{esc}$ is never zero. In fact the
distribution of escape times for slow dissipation will most likely
have a lower cutoff below which no dissipation is observed. Note from
Eq.(\ref{distau}) that $\gamma^{-1}$ is a measure of the lifetime of
the metastable states. The values obtained from the fits in
Fig.~\ref{eckhomFig7} are $\gamma^{-1}\approx 2$sec for the fast decay
and $\gamma^{-1}\approx 10^3$sec for the slow one.

This discussion supports the view that the experimentally observed
dissipation curves are averages over many events like the one shown in
Fig.~\ref{tanh} with an escape time distribution similar to
Fig.~\ref{taudistrib}.

\section{Conclusions}

We have presented a first study of the two dimensional planar $XY$
model in topologically non-trivial metastable states. These {\it
twisted} states represent a two dimensional superfluid under
non-equilibrium {\it persistent flow} conditions which satisfy the
superfluid velocity quantization condition, Eq.~(\ref{velocity}).
Transitions among these states change the topological quantum number,
the quantized flux, and represent the dissipation of flux quanta in a
flowing superfluid which is related to its critical velocity. The
properties of these transitions were studied in particular their
dependence on the geometry of the system and the superfluid
velocity. We showed that, in agreement with experimental results, the
metastable states have shorter lifetimes and, therefore, lower
critical velocities when the width of the system increases. Also in
agreement with experiments, we showed that dissipation is more rapid
the higher the initial superfluid velocity. The dissipation mechanism
was identified and studied: When two bands of spin differing by $2\pi$
deform and touch due to thermal fluctuations, a bound
vortex-antivortex pair of length $L_y$ is created. The confining force
pulls the vortex and antivortex towards each other thus zipping
together the two touching bands of spin into a single band.

Using this mechanism and a functional form for the escape time
distribution motivated by the numerical results, we calculated the
average superfluid velocity, ${\bar v}_s(t)$, and showed it to be in
excellent agreement both with our numerical simulations,
Fig.~\ref{dissip}, and with the experimental results of
reference~\cite{eckholm}, Figs.~\ref{eckhomFig6} and
~\ref{eckhomFig7}. This provides support for the view that the
dissipation observed experimentally is an average over several regions
with, possibly, different velocities where dissipation of flux quanta
takes place independently and at different times.

It is interesting that the two dimensional planar $XY$ model, with no
condensate and no linear dispersion for the low-lying excitations
still exhibits (meta)stable superfluid flow and agrees so well with
experimental results.

Finally we mention that the free energy landscape may be studied
directly using, for example, the Wang-Landau~\cite{wanglandau}
algorithm. Integer windings in the two dimensional $XY$ model were
addressed in~\cite{ggbleb}. Very recently~\cite{pdf}, different
topological sectors and the helicity modulus were studied for the four
dimensional compact $U(1)$ lattice gauge theory.

\acknowledgements 

The author acknowledges very helpful discussions with E. L Pollock,
B. Militzer, R. T. Scalettar, M. Troyer, Ph. de Forcrand and
T. Ramstad. He also thanks the Norwegian University of Science and
Technology, the Complex Group and Norsk Hydro for their hospitality
and generosity during a sabbatical stay.  This work was supported by
the NSF-CNRS cooperative grant \#12929.


\begin{thebibliography}{10}
\bibitem{lowangmom}
What this means will become clear shortly.
\bibitem{footnotevs} We use $v_s$ in units of $\hbar/m$, see
  Eq.~(\ref{Vs}).
\bibitem{note} Note that the exponent of the prefactor $L_y$ is always
  positive for $T< T_{KT}$ where superfluidty exists.
\end{thebibliography}
\end{document}